\documentclass{bmcart}
\usepackage{verbatim}
\usepackage[utf8]{inputenc} %unicode support
\usepackage{lipsum}
\usepackage{amsfonts,amsmath,amssymb,amsthm}
\usepackage{graphicx}
\usepackage{subfigure}
\usepackage[linesnumbered,ruled,vlined]{algorithm2e}
\usepackage{color,soul}
\usepackage[colorlinks=true,linkcolor=cyan,citecolor=blue]{hyperref}

\providecommand{\mb}[1]{\boldsymbol{#1}}

\newcommand{\EE}{\mathbb{E}}         
\newcommand{\bX}{\mb{X}}

\makeatletter
\AtBeginDocument{\let\hl\@firstofone}
\makeatother

\startlocaldefs
\endlocaldefs

\begin{document}

\begin{frontmatter}

\begin{fmbox}
\dochead{Research}

\title{
%Discovering a temporal change point \\ in a time series of organoid networks
Discovering a change point and piecewise linear structure in a time series of organoid networks \\via the iso-mirror
}

\author[
	addressref={aff1},
%	corref={aff1},
%	noteref={n1},
	email={tchen94@jh.edu}
]{\inits{TC}\fnm{Tianyi} \snm{Chen}}
\author[
	addressref={aff1},
%	corref={aff1},
%	noteref={n1},
	email={youngser@jhu.edu}
]{\inits{YP}\fnm{Youngser Park}}
\author[
	addressref={aff1},
%	corref={aff1},
%	noteref={n1},
	email={ali.saadeldin11@gmail.com}
]{\inits{ASE}\fnm{Ali Saad-Eldin}} 
\author[
	addressref={aff1},
%	corref={aff1},
%	noteref={n1},
	email={zlubber1@jhu.edu}
]{\inits{ZL}\fnm{Zachary Lubberts}}
\author[
	addressref={aff1},
%	corref={aff1},
%	noteref={n1},
	email={athpri@gmail.com}
]{\inits{AA}\fnm{Avanti Athreya}}
\author[
	addressref={aff1},
%	corref={aff1},
%	noteref={n1},
	email={bpedigo@jhu.edu}
]{\inits{BDP}\fnm{Benjamin D. Pedigo}}
\author[
	addressref={aff1},
%	corref={aff1},
%	noteref={n1},
	email={jovo@jhu.edu}
]{\inits{JTV}\fnm{Joshua T. Vogelstein}}
\author[
	addressref={aff2},
%	corref={aff1},
%	noteref={n1},
	email={fpuppo@health.ucsd.edu}
]{\inits{FP}\fnm{Francesca Puppo}}
\author[
	addressref={aff3},
%	corref={aff1},
%	noteref={n1},
	email={gsilva@health.ucsd.edu}
]{\inits{GAS}\fnm{Gabriel A. Silva}}
\author[
	addressref={aff2},
%	corref={aff1},
%	noteref={n1},
	email={muotri@health.ucsd.edu}
]{\inits{ARM}\fnm{Alysson R. Muotri}}
\author[
	addressref={aff4},
%	corref={aff1},
%	noteref={n1},
	email={weiwei.yang@microsoft.com}
]{\inits{WY}\fnm{Weiwei Yang}}
\author[
	addressref={aff4},
%	corref={aff1},
%	noteref={n1},
	email={chwh@microsoft.com}
]{\inits{CMW}\fnm{Christopher M. White}}
\author[
	addressref={aff1},
	corref={aff1},
%	noteref={n1},
	email={cep@jhu.edu}
]{\inits{CEP}\fnm{Carey E. Priebe}}
%\affil[1]{Johns Hopkins University}
%\affil[2]{University of California San Diego}
%\affil[3]{Microsoft Research}

\address[id=aff1]{%                           % unique id
  \orgname{Department of Applied Mathematics and Statistics, Johns Hopkins University}, % university, etc
%  \street{Waterloo Road},                     %
  %\postcode{}                                % post or zip code
  \city{Baltimore},                              % city
  \cny{USA}                                    % country
}
\address[id=aff2]{%
  \orgname{Department of Pediatrics, University of California, San Diego},
%  \street{D\"{u}sternbrooker Weg 20},
%  \postcode{24105}
  \city{San Diego},
  \cny{USA}
}
\address[id=aff3]{%
  \orgname{Department of Neurosciences, University of California, San Diego},
%  \street{D\"{u}sternbrooker Weg 20},
%  \postcode{24105}
  \city{San Diego},
  \cny{USA}
}
\address[id=aff4]{%
  \orgname{Microsoft Research},
%  \street{D\"{u}sternbrooker Weg 20},
%  \postcode{24105}
  \city{Seattle},
  \cny{USA}
}

\begin{artnotes}
%\note{Sample of title note}     % note to the article
%\note[id=n1]{Equal contributor} % note, connected to author
\end{artnotes}

\end{fmbox}% comment this for two column layout

%\setcounter{tocdepth}{2}
%\linespread{1.35} %regulate line spacing

\maketitle

\begin{abstractbox}

%\vskip0.1in
\begin{abstract}
Recent advancements have been made in the development of cell-based \textit{in-vitro} neuronal networks, or organoids. \hl{In order to better understand the network structure of these organoids, a super-selective algorithm has been proposed for inferring the effective connectivity networks from multi-electrode array data. In this paper, we apply a novel statistical method called spectral mirror estimation to the time series of inferred effective connectivity organoid networks.} This method produces a one-dimensional iso-mirror representation of the dynamics of the time series of the networks \hl{which exhibits a piecewise linear structure}. A classical change point algorithm is then applied to this representation, which successfully detects \hl{a change point 
%at 188 days (approximately 6 months). This change point coincides 
coinciding with the neuroscientifically significant time inhibitory neurons start appearing and the percentage of astrocytes increases dramatically}. This finding demonstrates the potential utility of applying the iso-mirror dynamic structure discovery method to inferred effective connectivity time series of organoid networks.
%can provide insight into identifying pattern changes in organoids.
\end{abstract}


% MSC classifications codes, if any
%\begin{keyword}[class=AMS]
%\kwd[Primary ]{}
%\kwd{}
%\kwd[; secondary ]{}
%\end{keyword}

\end{abstractbox}

\end{frontmatter}

\section*{Introduction}
Detecting structural changes in time series of networks is central to many modern network science applications.
However, due to the complexity of temporal network data and the myriad possible aspects for potential structural change, this problem can be daunting.
For discovering underlying dynamics in time series of networks, \cite{athreya2022discovering} proposes theory and methods for representing temporal network structure with a curve, or \hl{`mirror'}, in low dimensional Euclidean space, enabling the use of classical change point detection algorithms. In this paper, we estimate the mirror for a time series of brain organoid connectivity networks and subsequently identify change points.
%collected from UCSD. 
Because the mirror estimation method requires a 1-1 vertex correspondence for the networks across time, we first demonstrate that the putative 1-1 correspondence obtained directly from data collection is sufficiently accurate by comparing it to the vertex correspondence obtained via graph matching \cite{Vogelstein2015-cx}. Thence, mirror estimation and manifold learning recovers a 1-dimensional piecewise linear `iso-mirror' representation, with an evident slope change. By using the change point detection algorithm from \cite{bucher2021deviations} and break point estimation for piecewise linear models from \cite{muggeo2017interval}, we identify a change points neuroscientific significance coinciding with development stages.

We organize this paper as follows. In Section \textit{Organoids} 
we introduce our brain organoids data and the extraction of effective connectivity networks based on extracellular electrophysiology recordings. In Section \textit{Graph Matching} we define the graph matching problem and present the fast approximate assignment algorithm. In Section \textit{Discovering underlying dynamics in times series of networks} we introduce the mirror estimation method and relevant model assumptions. In the \textit{Results} section we apply these methods to the time series of organoid networks and present the graph matching results and the change point detection results. We conclude the paper with a discussion.

\section*{Organoids}
\label{sec:org}

%\textcolor{red}{
The brain organoids we consider are self-organizing structures composed of roughly 2.5 million neural cells. They are generated from human induced pluripotent stem cells (hiPSCs) \cite{muguruma2015self}. After growing for 6 weeks, they are plated in 8 wells of a multi-electrode array (MEA) plate (Axion Biosystem, Atlanta,GA, USA). MEA contains 64 low-impedance (0.04 MU) platinum microelectrodes with \hl{30 \textmu m of diameter spaced by 200 \textmu m}. Each well contains two or three organoids. Then, to characterize the functional development of the organoids, extracellular spontaneous electrical activity is recorded weekly using Maestro MEA system and AxIS Software Spontaneous Neural Configuration (Axion Biosystems)
with a \hl{0.1-Hz to 5-kHz} band-pass filter. Then spikes are detected with an adaptive
threshold crossing set to 5.5 times the standard deviation of the estimated noise for each
electrode. \hl{Each time series consists of five minutes of recorded neural activity across 10 months and the data are recorded irregularly -- not exactly once a week. Cortical organoids show low and sparse activity during the first 2 months with an average firing frequency of 0.5-Hz, then they start exhibiting highly synchronized and stereotypical network activity which transitions into 2-Hz and 3-Hz rhythmic activity by 4-6 months. At later stages (6 to 9 months), the activity includes high-rate spiking with peak of activity reaching a 20-Hz pace and highly complex bursting behaviors with cross-frequency coupling} \cite{puppo2021super}. \hl{A combination of principal component analysis and k-means clustering is used to spike sort the multi-unit activity from the 64 electrodes of each well. The average number of neurons detected in each well increases with the maturation of the organoids onto the electrodes, sometime reaching saturation after 6 months. Similar behavior is observed in all MEA wells. For example, the number of spike-sorted neurons in one well is detected as 122, 160, 189, 174, 190 at 2, 4, 6, 8 and 10 months. The other well has 81, 160, 174, 171, 170 neurons detected at 2, 4, 6, 8, and 10 months.}

%Five minutes of data were recorded. After processing, the spike-sorted activity of each well across 10 months is obtained by applying a standard protocol based on principal component analysis (PCA) and k-means clustering. \hl{Note the data are recorded irregularly -- not exactly once a week.} 

To infer the effective connections between neurons -- i.e., the adjacency matrices with neurons as vertices across time from the spike activity data -- the algorithm proposed in \cite{puppo2021super} is applied. This algorithm uses a super-selection rule
to individuate and discard correlation peaks corresponding to apparent and indirect interactions, and
reconstructs the effective connectivity of the network considering the remaining correlation
delays. \hl{ In the end, 45 effective connectivity networks on 127 vertices across 244 days are obtained. This is  our time series of organoid networks. }
%}

\section*{Graph Matching}
Given a time series of networks $G_1,\cdots,G_T$ on the same set of vertices $V$, a 1-1 vertex correspondence across all the networks facilitates joint spectral embedding, which is a key step in the mirror estimation method we will present in \hl{the next section, ``Discovering underlying dynamics in time series of networks}''. Such a vertex correspondence may be available a priori in labeled networks, or it can be inferred from unlabeled networks.
This inference problem -- the so-called graph matching problem -- is to find an alignment of vertices between two graphs such that the corresponding edge differences are minimized. We denote $A$, $B$ as two $n \times n$ adjacency matrices for two graphs with $n$ vertices each. Then the graph matching problem is to find the permutation matrix $P$ that maximizes the objective function
$$
f(A,B;P) = - ||A - P^{\top} B P||_F^2
$$
for $P \in \mathcal{P}$
where 
$\mathcal{P} = \{P \in \{0,1\}^{n\times n} : P^{\top}\mathbf{1}=P\mathbf{1}=\mathbf{1} \}$,
 $\mathbf{1} = (1,1,...1)^{\top}$, and $||\cdot||$ denotes the Frobenius norm.
%\begin{comment}
%\begin{equation}
%\begin{aligned}
%f(P) &= {\;||A - P^{\top} B P||_F^2}  \\
% P & \in \mathcal{P}  \\
% \mathcal{P}   = \{P  &\in \{0,1\}^{n\times n} : P^{\top}\mathbf{1}=P\mathbf{1}=\mathbf{1} \} \\
% \mathbf{1} & = (1,1,...1)^{\top}
%\end{aligned}
%\end{equation}
%\end{comment}
This formulation is equivalent to maximizing
\begin{equation*}
f(A,B;P) = trace (A P B^T P^T).
\end{equation*}
Because solving this optimization problem is combinatorically difficult, approximation algorithms have been proposed.
We use the Fast Approximate Quadratic (FAQ) assignment algorithm \cite{Vogelstein2015-cx} to obtain an approximate solution. FAQ iteratively finds a local solution to the relaxed problem -- expanding the constraint set to the convex hull of $\mathcal{P}$ -- and then projects the solution back to $\mathcal{P}$, and has been shown empirically to be competitive with or superior to other state-of-art methods. In Section Results we use FAQ to demonstrate that the given putative 1-1 correspondence for each pair of networks is close to the solution to the corresponding graph matching problem. 

\section*{Discovering underlying dynamics in time series of networks}
To discover the underlying dynamics in time series of networks, we use the model and method proposed in \cite{athreya2022discovering}. First we introduce the generative joint model for time series of networks. We consider $T$ networks, each containing $n$ vertices, with adjacency matrices $A_t$, $t \in \{1,2,...,T\}$.
In the model, each vertex is associated with a time varying $d$-dimensional latent vector. These vectors $X_t$ are each one realization from a stochastic process, called the latent position process (LPP) --
each $X_t$ is a $d$-dimensional random variable. 
For $n$ vertices, we generate $n$ i.i.d.\ samples from the LPP, which collection then forms the latent position matrices $\{\mathbf{X}_t\}$, where $\mathbf{X}_t \in \mathcal{R}^{n \times d}$ for $t \in \{1,2,...,T\}$.  
The connection probability between vertex $i$ and vertex $j$ at time $t$ is the inner product of the associated latent vectors at time $t$. That is $\EE(A_t)=\mathbf{X}_t\mathbf{X}^T_t$. Note that each network corresponds to a latent position random variable. Thus we can capture the distance between graphs using the corresponding random variables. We define the distance $$d_{MV}(X_t,X_{t'})=\min_{W \in \mathcal{O}^{d \times d}} \|\EE[(X_t-W X_{t'})(X_t-W X_{t'})^{\top}]\|^{1/2}_2,$$ \hl{where $\mathcal{O}^{d \times d}$ is the set of orthogonal transformation matrices with dimension $d$. When $X_t$ and $X_{t'}$ are centered, the $d_{MV}$ distance can be interpreted as the maximum directional variation for the random vector $X_t - WX_{t'}$, where $W$ is an orthogonal transformation used to align $X_t$ and $X_{t'}$.} 
We evaluate this distance for every pair of random variables in the LPP and obtain a $T \times T$ distance matrix $\mathcal{D}$. Then we apply classical multidimensional scaling (MDS) \cite{torgerson1952multidimensional} to $\mathcal{D}$ to get a low-dimensional Euclidean representation of underlying network structure, called the $\emph{mirror}$, $\{\psi(t)\}$.
In practice, the LPP is unknown and only network realizations $\{A_t\}$ are observed.
%$\{\hat{\mathbf{X}}_{t}\}$. 
For the $n \times n$ symmetrized adjacency matrix $A_t$, we use adjacency spectral embedding (ASE) \cite{Athreya2018-sa} to obtain
$\hat{\mathbf{X}}_{t} = U \Sigma^{1/2}$,
where the diagonal matrix $\Sigma$ contains the top $d$ eigenvalues of $A_t$ and $U$ contains the associated eigenvectors.  
%Besides ASE, we can also use graph encoder embedding\cite{shen2022one} or other joint embedding method such as OMNI \cite{levin2017central}.  
Then we use
$$
\hat{d}_{MV}(\hat{\bX}_{t},\hat{\bX}_{t'})=\min_{W} \frac{1}{\sqrt{n}} \|\hat{\bX}_{t}-\hat{\bX}_{t'}W\|^{1/2}_2
$$
to estimate the pairwise distance between networks, yielding $\hat{\mathcal{D}}$. Applying MDS to $\hat{\mathcal{D}}$ yields the mirror estimate $\{\hat{\psi}(t)\}$. 
%We call this procedure as Spectral MIRROR method. 
\hl{When the mirror exhibits a manifold structure, we can further simplify the change point detection problem by applying the manifold learning method isometric mapping (ISOMAP)} \cite{tenenbaum2000global} \hl{to $\{\hat{\psi}(t)\}$. This yields the \emph{iso-mirror}, which captures the geodesic distance along the mirror and preserves it in lower dimensions with Euclidean distance. Subsequent inference is then performed using the iso-mirror representation. For convenience, the iso-mirror representation will also be denoted as $\{\hat{\psi}(t)\}$.}

\section*{Results}
\subsection*{Time series of organoid networks data}
\begin{figure}[!ht] 
\centering
\includegraphics[width=0.8\textwidth]{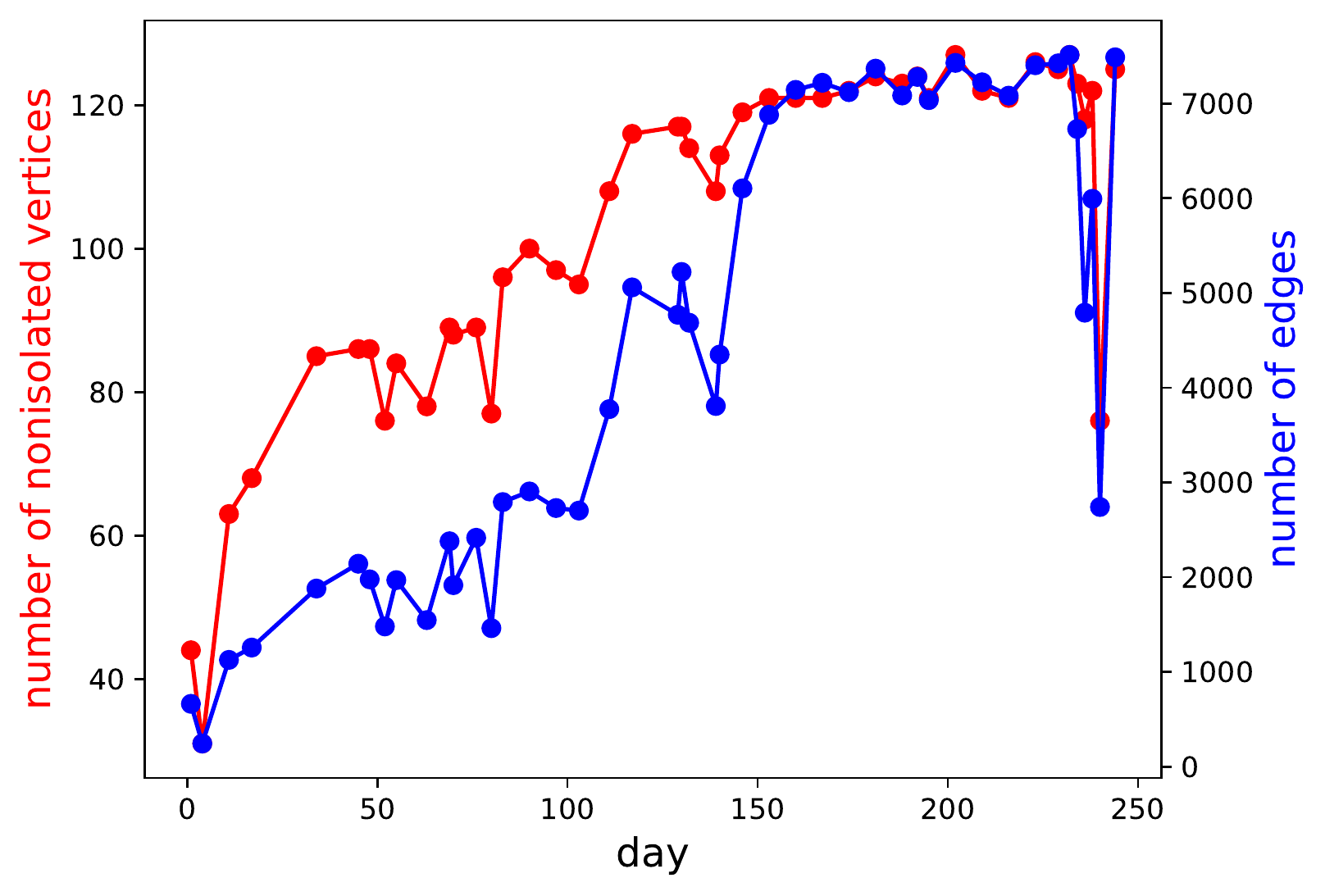} 
\caption {The number of non-isolated vertices and the number of edges for the graphs at each of the \hl{45 time stamps. The number of edges are counted after symmetrizing the directed graph.} }
\label{f1}
\end{figure}

\hl{All results in this section are based on data collected from well 8. For analogous results from well 5, please refer to the Appendix.} For well 8, the time series of organoid networks consists of \hl{ 45 time stamps $\{1,2,...,45\}$}; each time stamp corresponds an effective connectivity graph $G_t$ with adjacency matrix $A_t$. Each graph is directed, weighted,
%(scaled via pass-to-ranks?), 
and hollow. \hl{ We symmetrize the directed graphs, and use ranks in place of the raw edge weights. }
All graphs have the same vertex set $V= \{1,2,...n\}$ with \hl{$n = |V|=127$}, although some of the graphs contain isolated vertices.
%(degree equals to 0). 
See Figure \ref{f1}.

\subsection*{Putative 1-1 correspondence}
\begin{figure}[h!]
	\subfigure[\label{fig:fig2a}]{
            \includegraphics[width=0.8\textwidth]{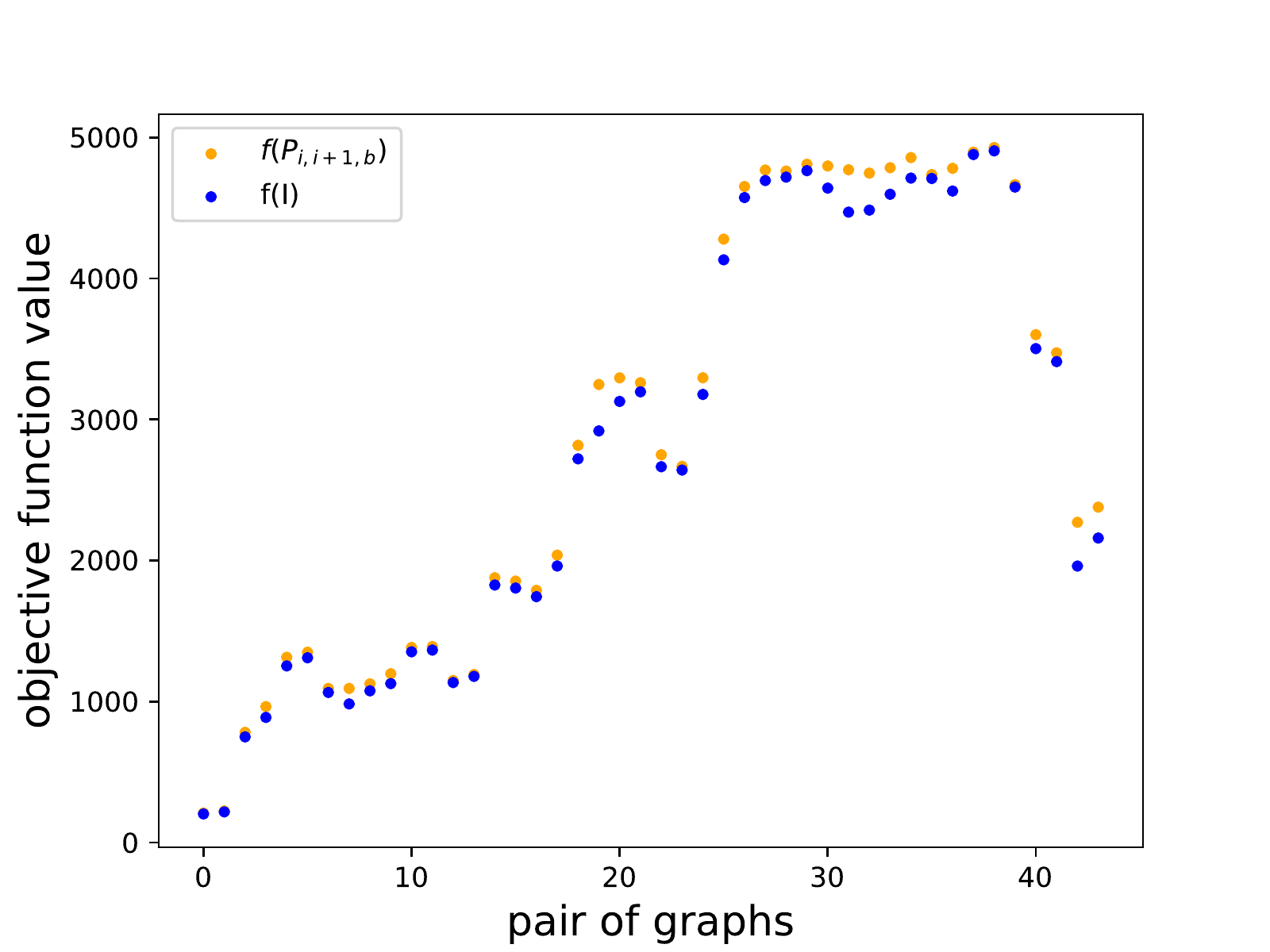}
	}
	\hfil
	\subfigure[\label{fig:rfig2b}]{
            \includegraphics[width=0.75\textwidth]{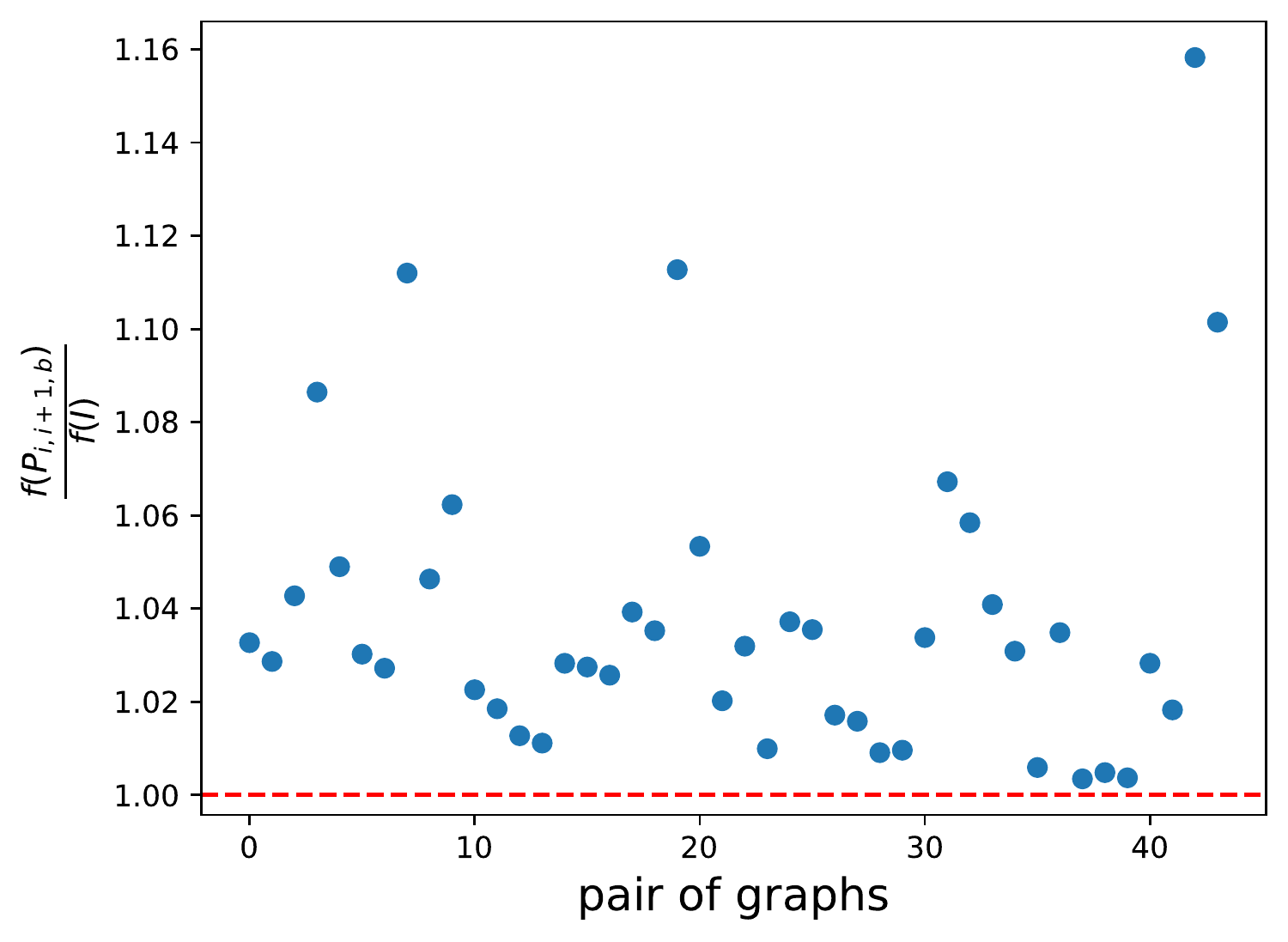}
	}
	\caption{Comparison of OFV using $I$ and $P_{i,i+1;b}$ for \hl{44} pairs of graphs, demonstrating that FAQ increases the OFV only slightly. (a): $f(I)$ and $f(P_{i,i+1;b})$. (b): $\frac{f(P_{i,i+1;b})}{f(I)}$.}
	\label{fig2}
\end{figure}

For this time series of organoid networks, 
%\textcolor{red}{
%physical locality alignment
inferred neuron location
%} 
gives a putative 1-1 vertex correspondence across the graphs. We \hl{assess} this correspondence via graph matching
%That is, we look for the permutation matrix $P \in \mathcal{P}$ which maximizes 
using the objective function value (OFV) $f(A,B;P)$.
%=trace(APB^TP^T)$.
The OFV for the putative 1-1 correspondence is given by $f(A,B;I)$ where $I$ is the identity matrix.
%Thus, essentially we are comparing whether OFV at the identity matrix (putative 1-1 correspondence) $f(A,B;I)$ is larger than the OFV at any other permutation matrix $f(A,B;P)$. However GMP is NP hard so instead we use Fast Approximate Quadratic programming(FAQ) \cite{Vogelstein2015-cx} to get an approximation solution. 
We denote the FAQ output for matching adjacency matrices $A$ and $B$ initialized at $C$ as $P_{A,B;C}$. 
Typically we choose the barycenter $b= \frac{\mathbf{1}\mathbf{1}^T}{n}$ as the initial point.
For all times \hl{$i \in \{1,2,...44\}$}, we consider $A_i$, $A_{i+1}$ and FAQ yields $P_{A_i,A_{i+1};b}$ (denoted $P_{i,i+1;b}$ for short). 
In Figure \ref{fig2} we compare $f(I) = f(A_i,A_{i+1};I)$ and $f(P_{i,i+1;b}) = f(A_i,A_{i+1};P_{i,i+1;b})$.
Although $f(P_{i,i+1;b})$ is always larger than $f(I)$, the two OFVs are close to each other for all time stamps, indicating that the putative 1-1 correspondence is close to FAQ's solution. 

\begin{figure}[h!]
	\subfigure[\label{fig:fig3a}]{
            \includegraphics[width=0.8\textwidth]{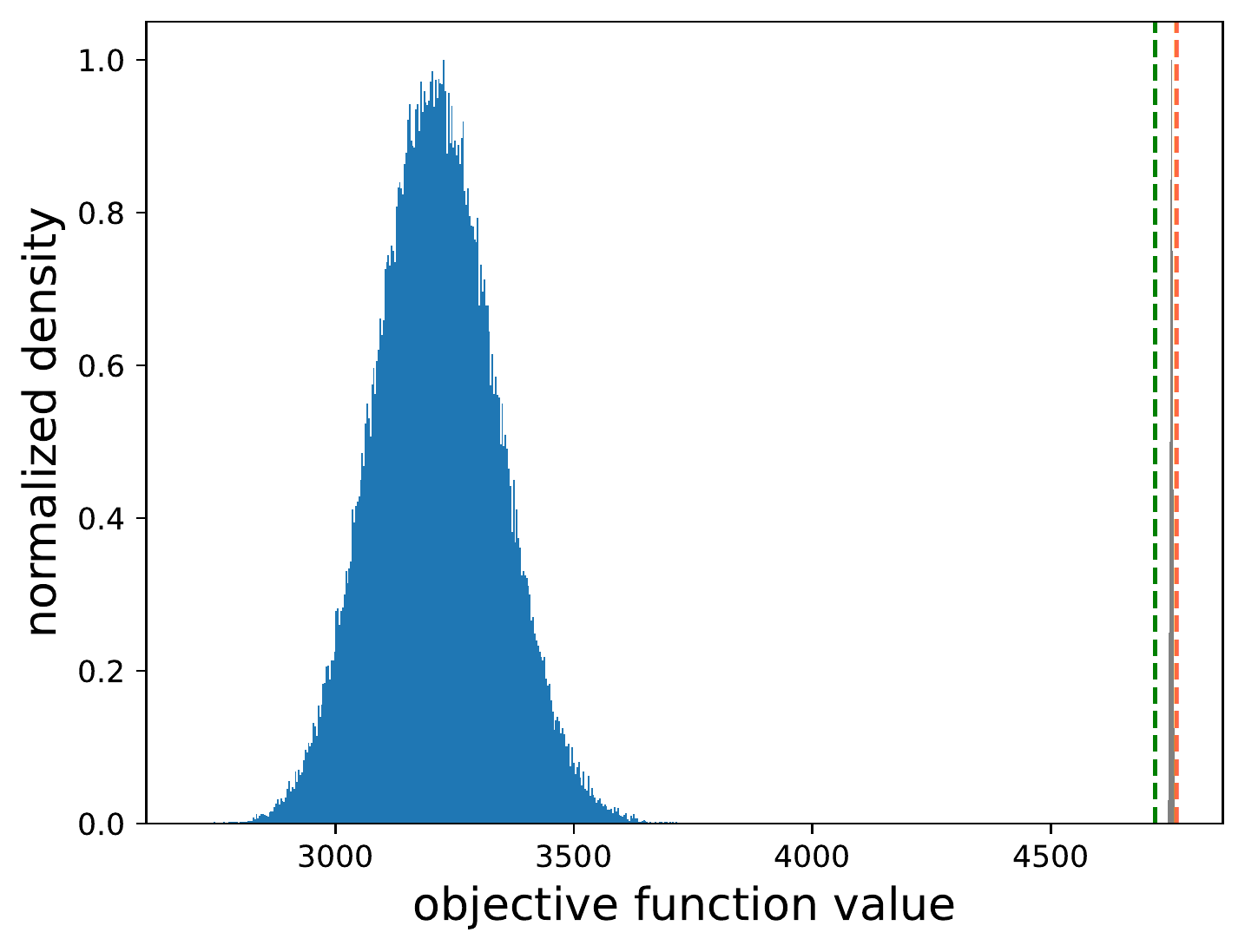}
	}
	\hfil
	\subfigure[\label{fig:fig3b}]{
             \includegraphics[width=0.8\textwidth]{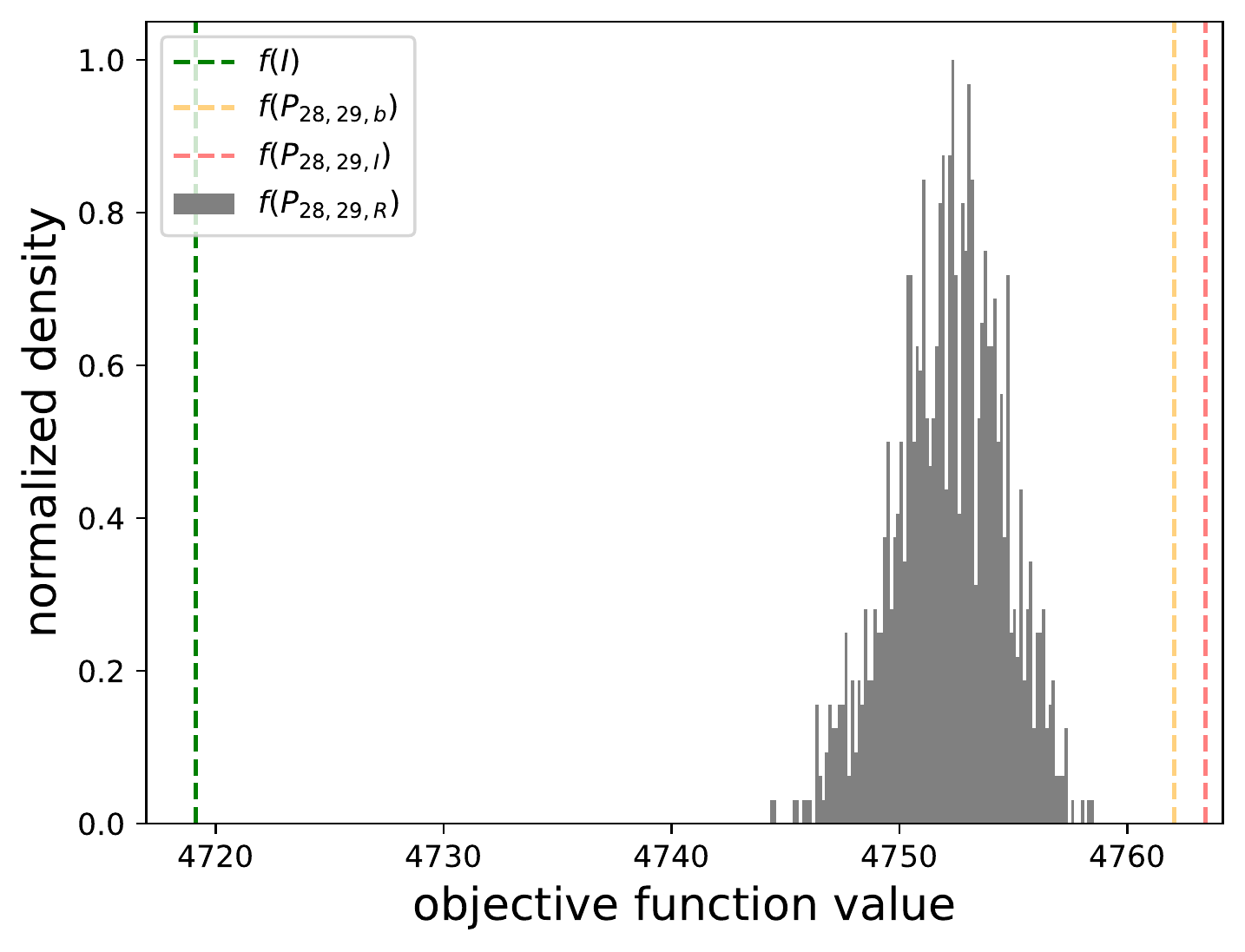}
	}
	\caption{Matching $A_{28}$ and $A_{29}$. Comparison of OFV for using $I$,$R$, $P_{28,29;b}$, $P_{28,29;I}$ and $P_{28,29;R}$. (a): Histogram demonstrating that OFV for $R$ is not nearly as good as for the others. (b): Enlargement of the far right portion of the top figure, demonstrating that FAQ output at different initial points $I$, $b$ and $R$ improves the OFV only slightly compared to the putative 1-1, $I$. }
	\label{fig3}
\end{figure}

To further asses the putative 1-1 correspondence, we consider a specific pair of adjacency matrices: $A_{28}$, $A_{29}$. We uniformly generate 100,000 random permutation matrices $R$ and evaluate $f(R) = f(A_{28},A_{29};R)$. We plot the histogram in Figure \ref{fig3}. We also indicate $f(A_{28},A_{29};I)$, $f(A_{28},A_{29};P_{28,29;I})$, $f(A_{28},A_{29};P_{28,29;b})$ and $f(A_{28},A_{29};P_{28,29;R})$, where $R$ are 100 randomly drawn permutation matrices. We see that the putative 1-1 correspondence performs better than all 100,000 instantiations of $f(R)$ and is close to $P_{28,29;b}$, $P_{28,29;I}$ and $P_{28,29;R}$. Thus we conclude that the putative 1-1 correspondence is sufficiently accurate, and we will proceed apace for mirror estimation and change point detection.

\subsection*{Change point detection}
For the \hl{45} graphs, time stamps are from 1 to 244, in days. We choose time stamps in [150,230] to avoid growth and death regimes.
%this period contains 13 graphs. For demonstration reason we only 

\begin{figure}[ht!] 
\centering
\includegraphics[width=.75\textwidth]{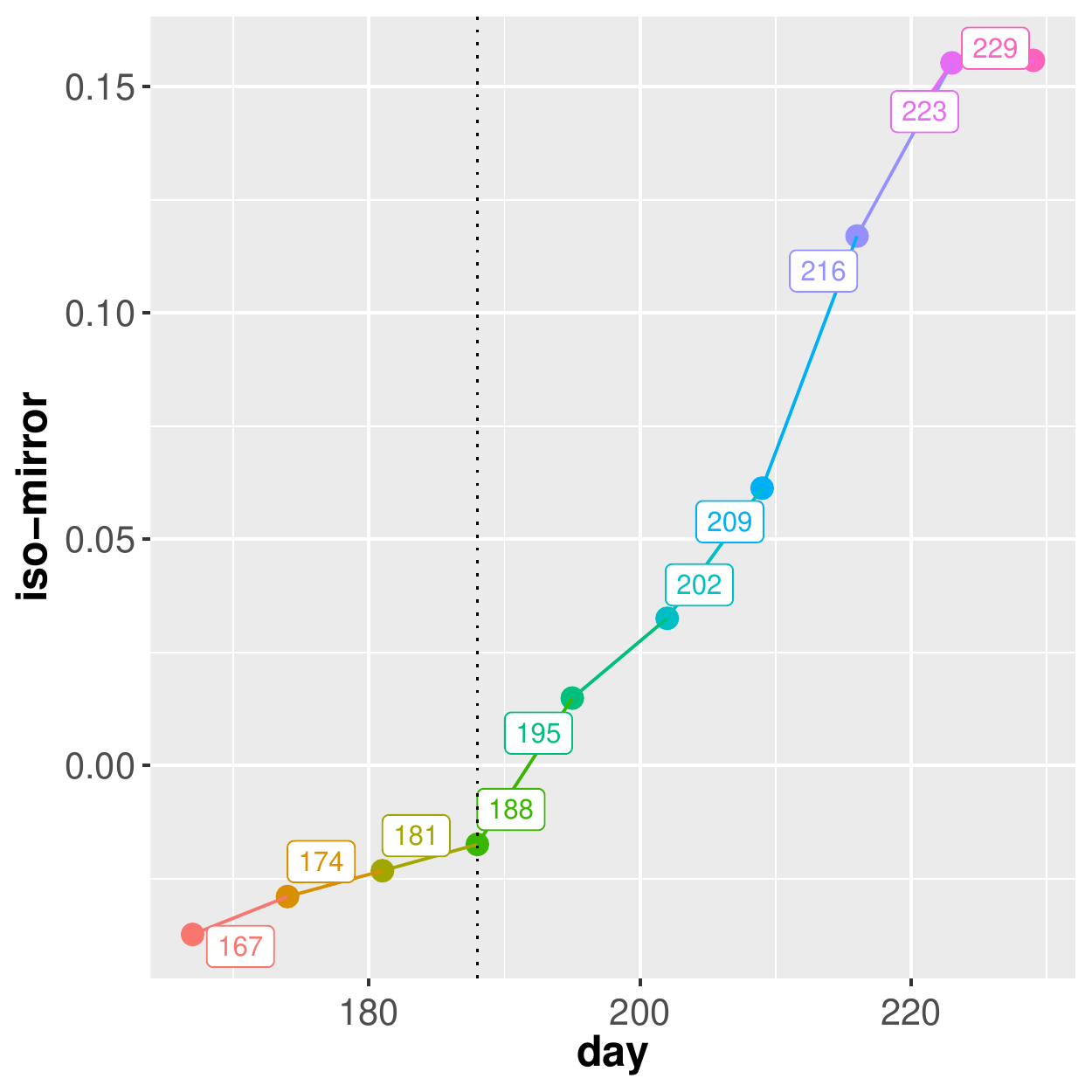}
\caption {Discovering a change point in time series of inferred effective connectivity organoid networks via the iso-mirror. The x-axis is time stamp, in days. The y-axis is the ISOMAP representation of the estimated mirror $\{\hat{\psi}_t\}$. 
%For illustration purpose, only 10 points are shown. 
This indicates a change in the network dynamics at day 188.}
\label{fig9}
\end{figure}

%and we remove the graphs with time stamps 153, 160 and 192.
For these graphs, we find the largest common connected component, which contains 112 vertices. The average number of edges for the largest common connected component is approximately 6130. We use the putative 1-1 vertex correspondence across time. We apply our mirror estimation method to this time series of networks, and ISOMAP manifold learning yields the 1-dimensional representation of the dynamics $\{\hat{\psi}_t\}$. \hl{We choose 10 time stamps
%, $t \in \{1,2,...10\}$, 
shown in Figure} \ref{fig9}. As we see, the representation is approximately piecewise linear with an evident change of slope at $t=4$, day 188.  

Assuming the true underlying 1-dimensional representation $\psi(t), t \in [0,T]$, is piecewise linear and continuous, it is natural to define the change point $t^*$ as the point when the slope changes. If we assume there is only one change point, then we can write
$$
\psi(t)=\beta_0+\beta_1t+\beta(t-t^*)I(t>t^*).
$$
Both the change point detection algorithm from \cite{bucher2021deviations} and the break point estimation for piecewise linear models from \cite{muggeo2017interval} yield an estimated change point $\hat{t^*}=4$, day 188,
%\textcolor{red}{
which  \hl{coincides with neuroscientifically significant developmental changes
--
inhibitory neurons start appearing and the percentage of astrocytes increases dramatically
--}
as described in \cite{trujillo2019complex}.%}.
\hl{ Note that the emergence of astrocytes does not happen at once but builds over time,
so there is no one precise date for the change point and detection of a time coinciding with this change can be but suggestive.}

%  I think we should mention somewhere in the discussion section about how the change point detected (188, 195, 209 day) coincides with development stages of the organoids describe in this paper , for example, the inhibitory neurons were observed around 6month etc.

%Applying it on our data, it yields figure \ref{fig11}. The break point estimation is day 190.2152, the closest time stamp to it is day 188, showing the results from two algorithms coincide.  

%\begin{figure}[h!] 
%\centering
%\includegraphics[width=.6\textwidth]{f11.pdf} 
%\caption {The break point estimation result is 190.2152, we choose the closest time stamp as the change point estimation, day 188. The initial guess of the break point is 180.  }
%\label{fig11}
%\end{figure} 

%\newpage

\section*{Conclusion}

Reconstruction of effective connectivity networks of electrophysiologically active brain organoids reflect their structural (increasing number of neurons (nodes) and connections (edges)) and electrical development over time, as previously demonstrated in \cite{trujillo2019complex}.

By applying the spectral mirror estimation method to the time series of organoid networks, we obtain a 1-dimensional iso-mirror representation of dynamic inferred effective connectivity organoid networks. Two change point detection algorithms successfully detect 
%the neuroscientifically significant 
\hl{a} change at day 188. 
At \hl{approximately} 188 days (\textasciitilde 6 months), cortical organoids start showing inhibitory neurons and the percentage of astrocytes increases from 5\% to 30-40\% \cite{trujillo2019complex} \hl{resulting in added complexity in the activity and network distribution of brain organoids.
}

\hl{There are several change point detection algorithms available for analyzing time series of graphs, including the one proposed in} \cite{wang2021optimal}. \hl{However, it is important to note that the spectral mirror estimation method used in our study is not restricted to change point detection alone. Instead, it provides a low-dimensional (in our case, one-dimensional Euclidean) representation of network dynamics, enabling us to visualize network evolution. As illustrated in Figure} \ref{fig9}, \hl{we observe piecewise linear structure and apply segment regression to identify a significant increase in slope after day 188. This suggests that organoid graphs drift continuously over time, but their rate of drift accelerates significantly after day 188. Our method is preferred in this regard as it provides us with more than just one change point. Additionally, since the iso-mirror represents the underlying LPP, the detected change point in the iso-mirror reflects a fundamental change in the underlying generative LPP of the time series of networks, which may not necessarily correspond to any specific network measure. Furthermore, the spectral mirror method can be applied to other dynamic graphs that meet our model assumptions, namely that the time series of networks are generated from a time series of latent position random networks whose vertices have independent, identically distributed latent positions given by an LPP.}

Future work includes addressing two major theoretical issues of note, to make the change point inference formally principled.
First, the theory in \cite{athreya2022discovering} requires a known 1-1 vertex correspondence across time. It remains to study the effect of errors in this correspondence, such as those inherent in our putative 1-1 correspondence deemed sufficiently accurate for practical purposes.
In addition, it remains to study the entry-wise behavior of the error term $\epsilon(t)=\hat{\psi}(t)-\psi(t)$. For example: in \cite{bucher2021deviations} the proof of consistency of the change point estimator requires the error process to be stationary; in  \cite{muggeo2017interval} the $\epsilon(t)$ are assumed to be i.i.d.\ normal to construct a confidence interval for $\hat{t^*}$. For now, \cite{athreya2022discovering} has shown that $\hat{\psi}(t)$ converges to $\psi(t)$ in Frobenius norm, that is $\sum_{t=1}^T(\epsilon(t))^2 \to 0$ with high probability, which is insufficient to conclude normality or stationarity. What's more, this convergence result is for the mirror rather than the iso-mirror. As for whether the same conclusion can be extended to the iso-mirror, further investigation is needed.
%not enough to conclude the distribution or ``stationaryness" of $\epsilon(t)$. 

% The bibliography file has a relatively recent copy of all neurodata pubs.

\section*{Appendix}

\subsection*{Time series of organoid networks data for well 5}
\begin{figure}[!ht] 
\centering
\includegraphics[width=.68\textwidth]{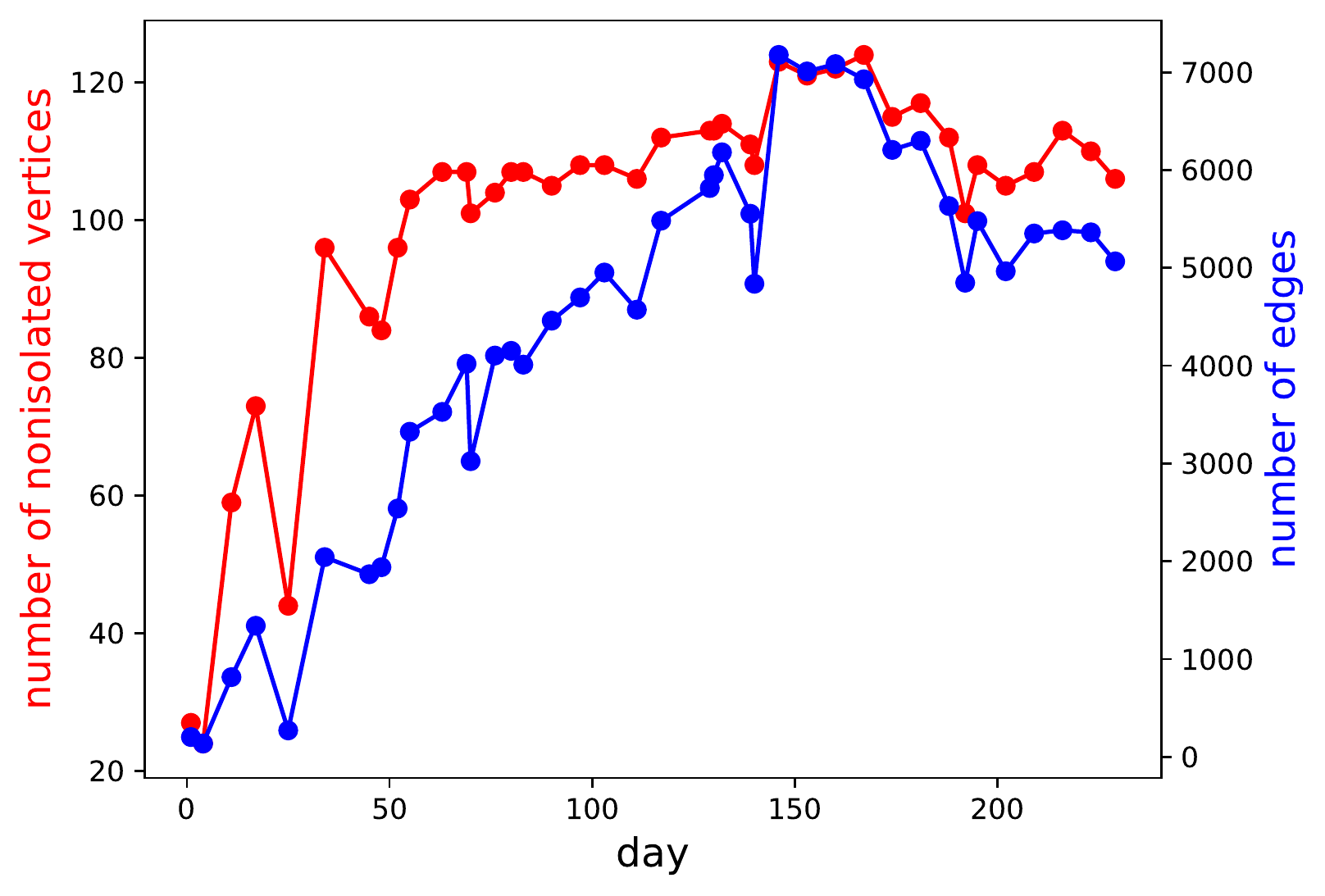} 
\caption {The number of non-isolated vertices and the number of edges for the graphs at each of the 40 time stamps for well 5. The number of edges are counted after symmetrizing the directed graph.}
\label{well5f1}
\end{figure}

For well 5, there are 40 graphs with time stamps [1,229] and all of them have the same vertex set with $|V|=128$, although some of the graphs contain isolated vertices. See figure \ref{well5f1}. Each graph is directed, weighted, and hollow. We symmetrize the directed graphs, and use ranks in place of the raw edge weights.

\subsection*{Putative 1-1 correspondence for well 5}

\begin{figure}[h!]
	\subfigure[\label{well5fig2a}]{
            \includegraphics[width=0.8\textwidth]{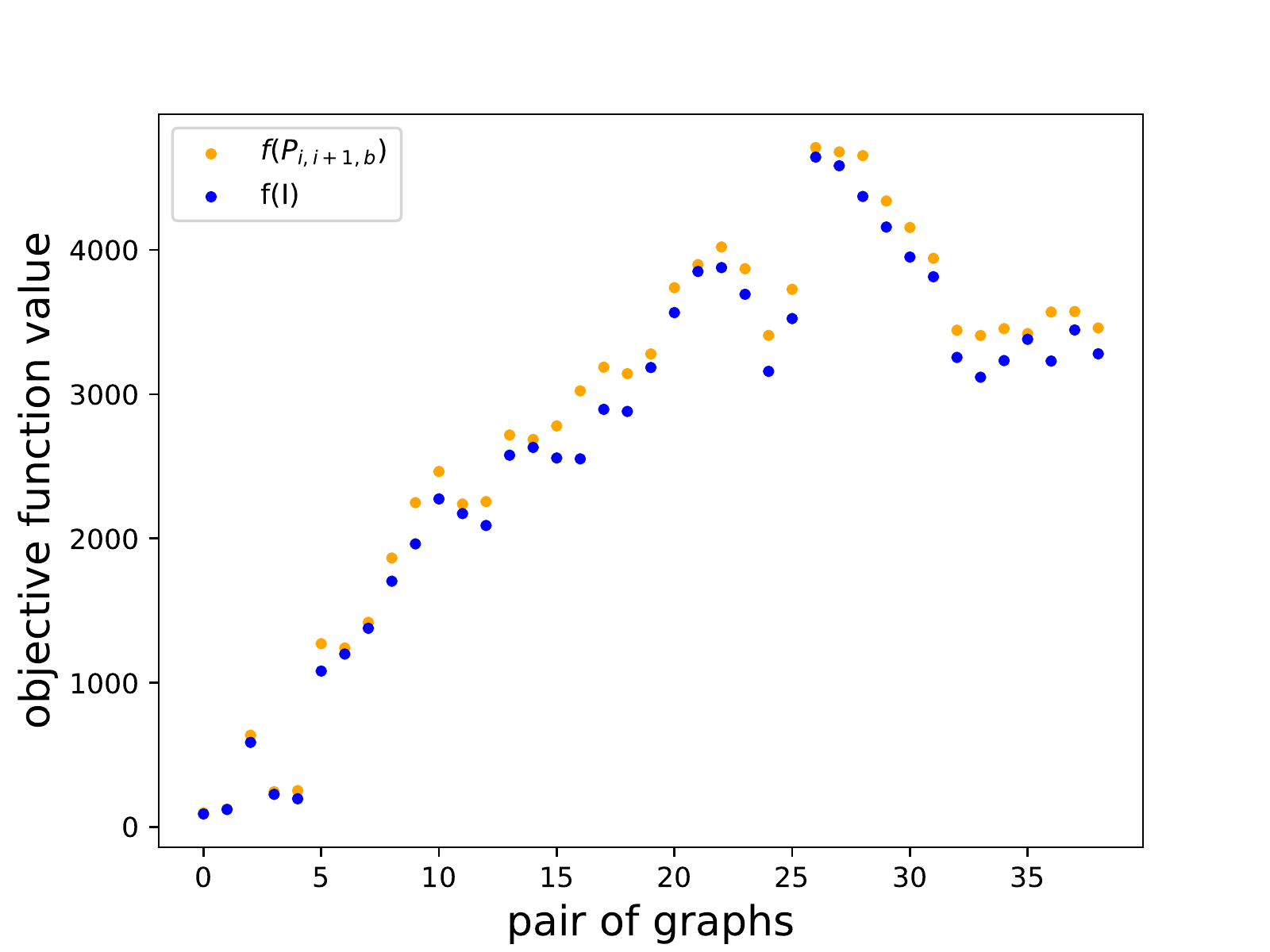}
	}
	\hfil
	\subfigure[\label{well5fig2b}]{
            \includegraphics[width=0.75\textwidth]{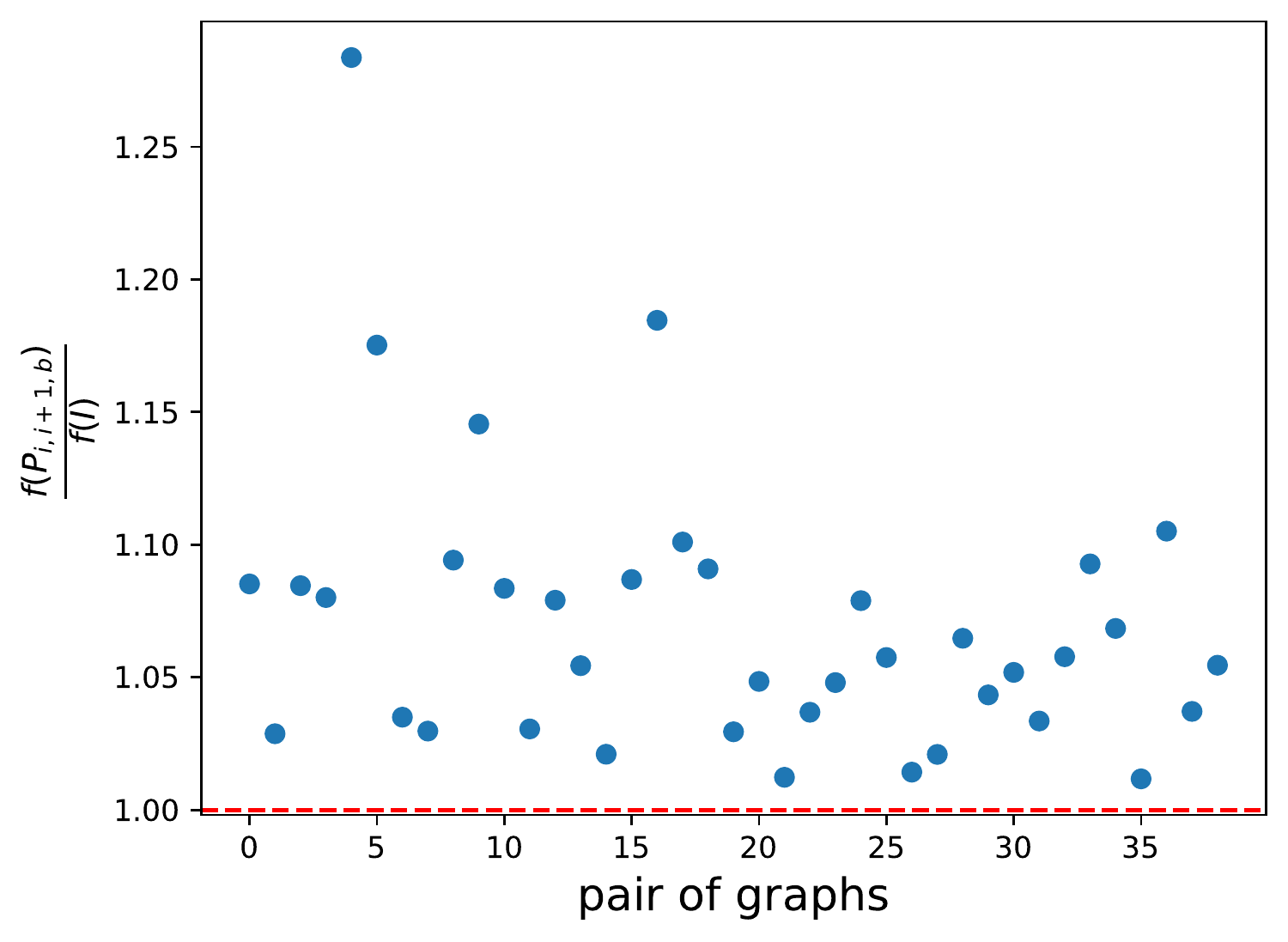}
	}
	\caption{Comparison of OFV using $I$ and $P_{i,i+1;b}$ for 39 pairs of graphs, demonstrating that FAQ increases the OFV only slightly. (a): $f(I)$ and $f(P_{i,i+1;b})$. (b): $\frac{f(P_{i,i+1;b})}{f(I)}$.}
	\label{well5fig2}
\end{figure}

We assess the putative 1-1 correspondence exactly the same way as in the paper and Figure \ref{well5fig2} indicates that the putative 1-1 correspondence is close to FAQ's solution.  

\begin{figure}[ht!]
     \subfigure{
         \centering
         \includegraphics[width=0.8\textwidth]{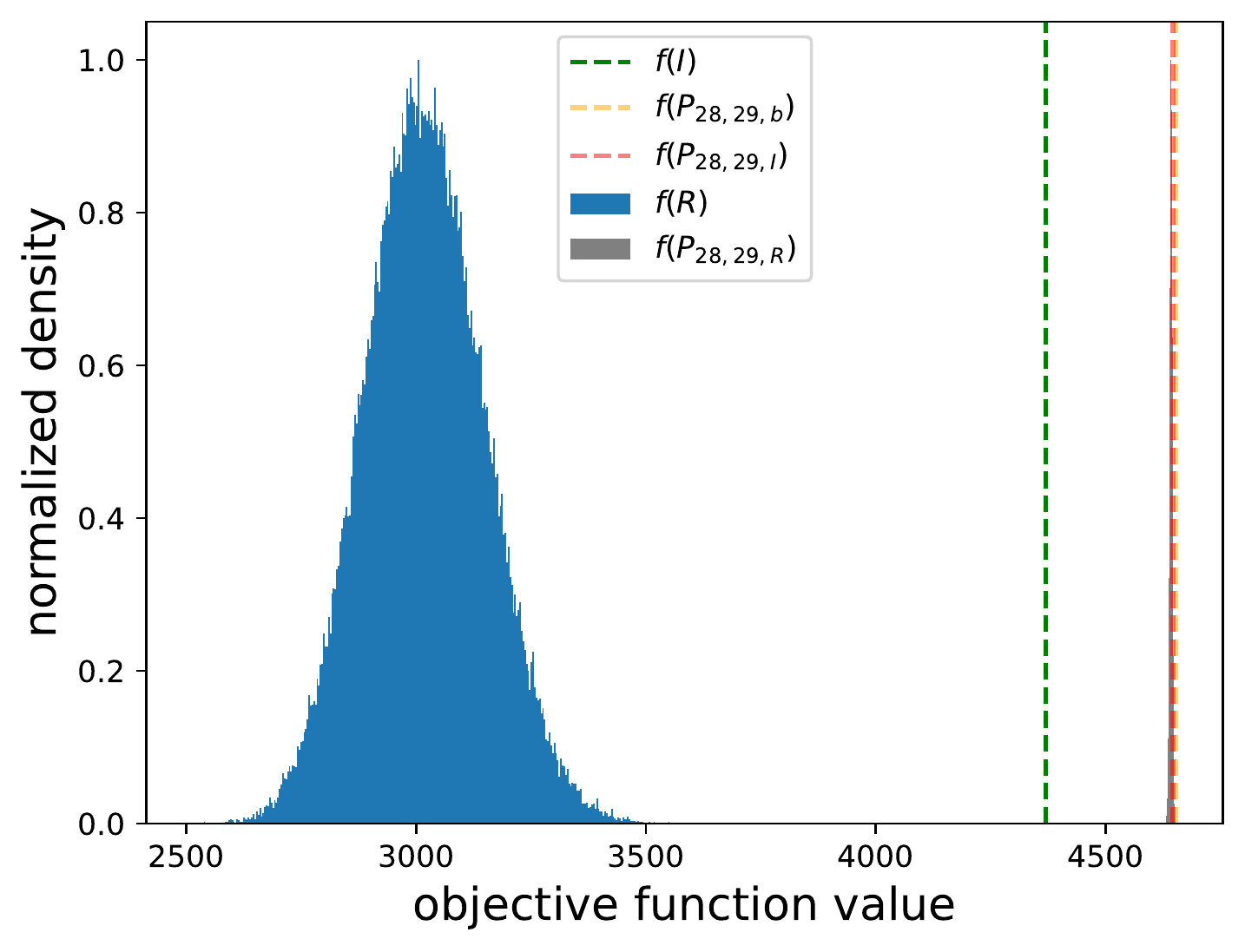}
         %\caption{Left part of the histogram}
         }
     \hfill
     \subfigure{
         \includegraphics[width=0.8\textwidth]{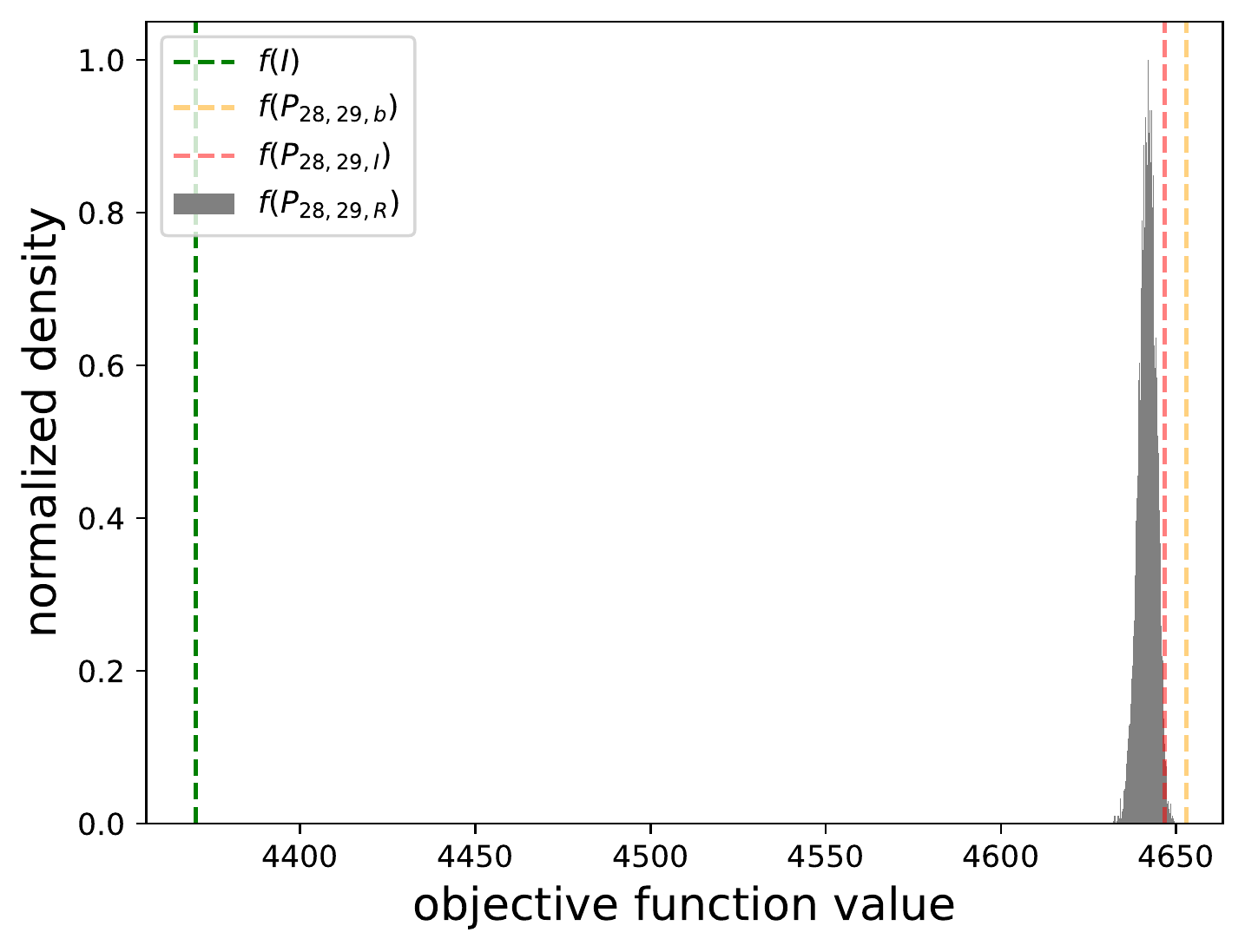}
         %\caption{Right part of the hisotgram}
         }
     \hfill
        \centering
        \caption{Matching $A_{28}$ and $A_{29}$. Comparison of OFV for using $I$,$R$, $P_{28,29;b}$, $P_{28,29;I}$ and $P_{28,29;R}$. Left: Histogram demonstrating that OFV for $R$ is not nearly as good as for the others. Right: Enlargement of the far right portion of the left figure, demonstrating that FAQ output at different initial points $I$, $b$ and $R$ improves the OFV only slightly compared to the putative 1-1, $I$.    }
        \label{well5fig3}
\end{figure}

We also consider a specific pair of adjacency matrices: $A_{28}$, $A_{29}$ to assess the 1-1 putative correspondence. See figure \ref{well5fig3}. The result is similar to well 8 and we conclude the 1-1 putative correspondence is accurate enough. 

\subsection*{Change point detection for well 5}
\begin{figure}[!ht] 
\centering
\includegraphics[width=.75\textwidth]{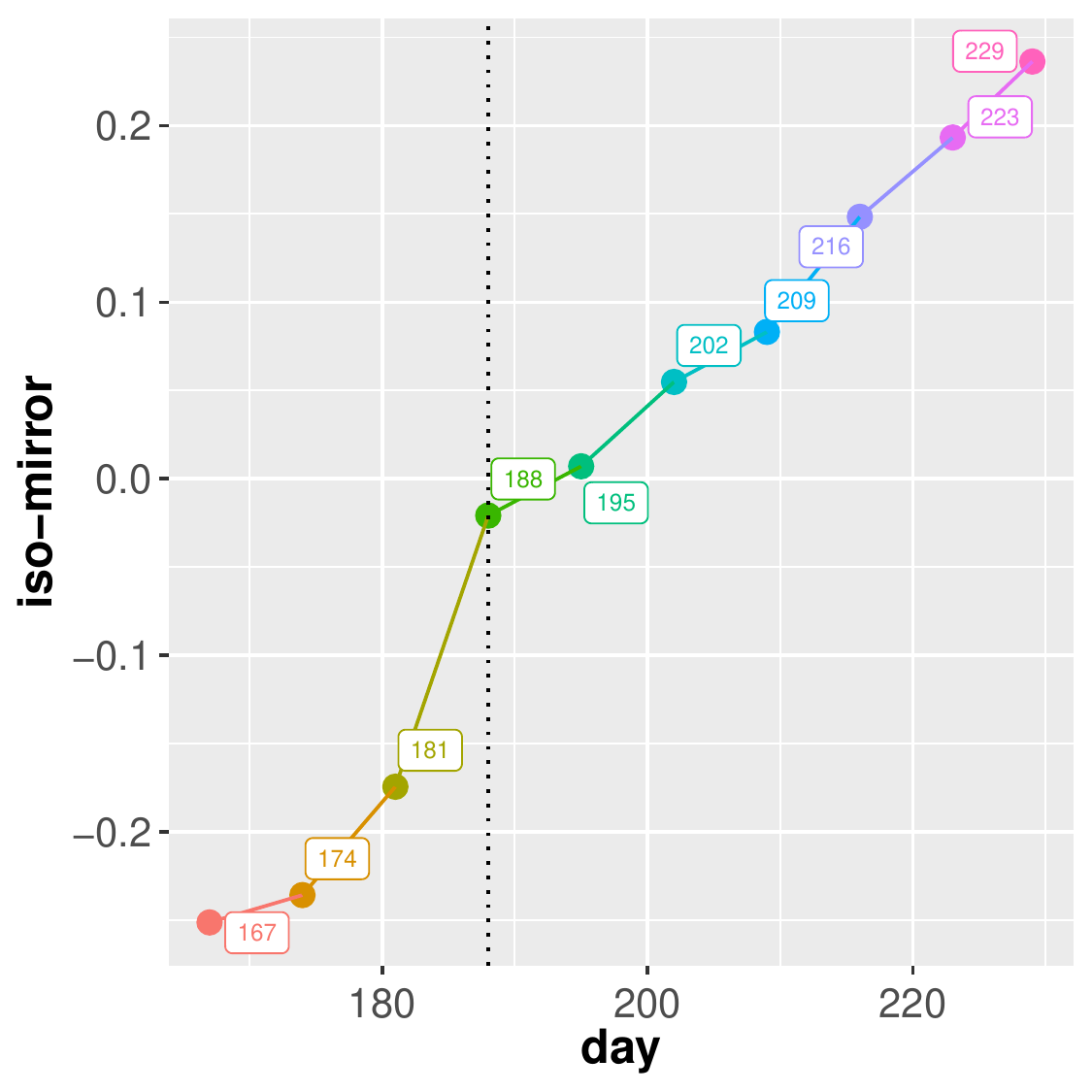} 
\caption {For well 5, iso-mirror result. }
\label{well5f5}
\end{figure}

For the 40 graphs, time stamps are from 1 to 244, in days. We choose the same time stamps in [150,230] as in the paper. We find the largest common connected component for these graphs. We use the putative 1-1 vertex correspondence across time. We apply our mirror estimation method to this time series of networks, and ISOMAP manifold learning yields the 1-dimensional representation of the dynamics $\{\hat{\psi}_t\}$. We choose the same 10 time stamps as in the paper 
%, $t \in \{1,2,...10\}$, 
shown in Figure \ref{well5f5}. As we see, the representation is approximately piecewise linear. The break point estimation for piecewise linear models from \cite{muggeo2017interval} yield an estimated change point $\hat{t^*}=4$, day 188.    

%\clearpage

\begin{backmatter}

\section*{Ethical interest}
not applicable

\section*{Competing interests}
The authors declare that they have no competing interests.

\section*{Abbreviations}%% if any
\textbf{MEA}: Multi-Electrode Array \\
\textbf{PCA}: Principal Component Analysis \\
\textbf{FAQ}: Fast Approximate Quadratic \\
\textbf{LPP}: Latent Position Process \\
\textbf{MDS}: Multidimensional Scaling \\
\hl{\textbf{ISOMAP}: Isometric Mapping} \\
\textbf{ASE}: Adjacency Spectral Embedding \\
\textbf{OFV}: Objective Function Value \\

\section*{Funding}%% if any
TC's work is partially supported by the Johns Hopkins Mathematical Institute for Data Science (MINDS) Data Science Fellowship and Azure sponsorship credits granted by Microsoft’s AI for Good Research Lab.

\section*{Availability of data and materials}%% if any
\hl{The organoid data and code are available at} \url{https://github.com/youngser/organoid}.

\section*{Authors' contributions}
TC, ZL, AA, CEP developed the theory. ARM, FP, GAS provided the data. TC, YP, AS, BPD designed \& implemented the methods. TC, YP conducted the experiments. TC, YP, CEP wrote the manuscript.  WY, CW, JTV, CEP guided the whole process. All authors read and approved the manuscript.

\bibliographystyle{bmc-mathphys} % Style BST file (bmc-mathphys, vancouver, spbasic).
\bibliography{ans.bib}

\end{backmatter}

\end{document}